\documentclass{nature}

\usepackage[dvipdf]{graphicx}
\usepackage{amssymb}
\usepackage{color}
\usepackage{rotating}
\usepackage{caption}

\newcommand \apj{\emph{Astrophys. J.}}
\newcommand \apjl{\emph{Astrophys. J. Lett.}}

\newcommand \aj{Astron. J.}
\newcommand \aap{\emph{Astron. Astrophys.}}

\newcommand \araa{Annual Rev. Astron. Astrophys.}
\newcommand \mnras{\emph{Mon. Not. R. Astron. Soc.}}
\newcommand \prc{\emph{Phys. Rev. C}}

\newcommand \nat{Nature}
\newcommand \procspie{Proc. SPIE}

\newcommand{\Ka}{K$\alpha$}

\newcommand{\NH}{$N_{\rm H}$}
\newcommand{\Msun}{$M_{\odot}$}
\newcommand{\Mch}{$M_{\rm Ch}$}

\title{Solar Abundance Ratios of the Iron-Peak Elements in the Perseus Cluster}

\author{Hitomi Collaboration (2017) \ --- \ {\it Nature}, 551, 478 (doi:10.1038/nature24301)}

\begin{document}

\maketitle


\medskip

\begin{abstract}
The metal abundance of the hot plasma that permeates galaxy clusters represents the accumulation 
of heavy elements produced by billions of supernovae\cite{Mushotzky96}. 
Therefore, X-ray spectroscopy of the intracluster medium provides an opportunity to investigate the nature of 
supernova explosions integrated over cosmic time. In particular, the abundance of the iron-peak elements 
(chromium, manganese, iron and nickel) is key to understanding how the progenitors of typical type Ia 
supernovae evolve and explode\cite{Finoguenov02,Mernier16b,Maeda10a,Seitenzahl13b,Yamaguchi15}. 
Recent X-ray studies of the intracluster medium found that the abundance ratios of these elements 
differ substantially from those seen in the Sun\cite{Dupke00,Gastaldello04,dePlaa07,DeGrandi09,Mernier16a}, 
suggesting differences between the nature of type Ia supernovae in the clusters and in the Milky Way. 
However, because the K-shell transition lines of chromium and manganese are weak and those of iron 
and nickel are very close in photon energy, high-resolution spectroscopy is required for an accurate 
determination of the abundances of these elements. 
Here we report observations of the Perseus cluster, with statistically significant detections of the resonance 
emission from chromium, manganese and nickel. Our measurements, combined with the latest atomic models, 
reveal that these elements have near-solar abundance ratios with respect to iron, in contrast to previous claims.
Comparison between our results and modern nucleosynthesis calculations\cite{Seitenzahl13a,Pakmor12,Woosley11} 
disfavours the hypothesis that type Ia supernova progenitors are exclusively white dwarfs with masses well below 
the Chandrasekhar limit (about 1.4 times the mass of the Sun). The observed abundance pattern of the iron-peak 
elements can be explained by taking into account a combination of near- and sub-Chandrasekhar-mass type Ia 
supernova systems, adding to the mounting evidence that both progenitor types make a substantial contribution 
to cosmic chemical enrichment\cite{Seitenzahl13b,Scalzo14,Blondin17}.
\end{abstract}

\medskip


The Soft X-ray Spectrometer (SXS) on board Hitomi achieved unprecedented spectral resolution in orbit 
($\Delta E \approx 5$\,eV in the 2--10\,keV band)\cite{Hitomi16}. Fig.\,1 shows the SXS spectrum of 
the Perseus Cluster core ($r$ $\lesssim$ $2'$ $\approx$ 40\,kpc) in the 1.8--9.0\,keV band. 
This was obtained from the same series of observations as our previous work that constrained 
turbulent velocities in the intracluster medium (ICM)\cite{Hitomi16}, 
but with 25\% more exposure totaling 290\,ks. The refined calibration of the telescope effective 
area and the SXS aperture window transmission now allows the first flux measurement of each individual 
line in the 1.8--9.0 keV band, encompassing the H- and He-like transitions from Si through Ni.

The excellent performance of the SXS also makes possible the detection of weak resonance lines 
from He-like Cr, Mn, and Ni, with statistical significance of 6$\sigma$, 4$\sigma$, and 12$\sigma$, 
respectively (Fig.\,1b and 1c). 
Measurements of these line fluxes in celestial sources have been extremely challenging with  
traditional non-dispersive X-ray detectors (e.g., charge coupled devices, or CCDs), 
because such weak features readily 
blend into the bremsstrahlung continuum under lower spectral resolution and 
the Ni~{\footnotesize XXVII} He$\alpha$ and Fe~{\footnotesize XXV} 
He$\beta$ lines cannot be resolved (see Fig.\,1c).

The hot ICM, confined in the deep cluster gravitational potential well, 
contains the dominant fraction ($\sim$80\%) of metals in the cluster\cite{Mushotzky96}. 
Among these, the Fe-peak elements (Cr, Mn, Fe, and Ni) are thought to be predominantly created by 
type Ia supernovae (SNe Ia) occurring over a cosmological time period\cite{Loewenstein96,Matsushita13}. 
Therefore, the abundance of these elements provides crucial information about the integrated 
SN Ia nucleosynthesis and its relevant physics.

Despite the importance of SNe Ia as distance indicators in cosmology\cite{Riess98,Perlmutter99}, 
many of their fundamental aspects remain elusive. One important open question is whether the mass 
of an exploding white dwarf (WD) is close to the Chandrasekhar limit ($M_{\rm Ch} \approx 1.4\,M_{\odot}$), 
regardless of whether it originates from a single WD accreting mass from a non-degenerate 
companion\cite{Whelan73} or a violent merger of two WDs\cite{Webbink84}.
Recent hydrodynamical simulations show that both so-called delayed-detonation explosions of 
near-\Mch\ WDs\cite{Maeda10a,Seitenzahl13a} and full detonations of 
sub-\Mch\ WDs\cite{Sim10,Woosley11,Pakmor12} can reproduce the observed properties 
(such as optical light curves and spectra) of SNe Ia. Therefore, it is difficult to distinguish 
the two scenarios from optical observations of individual explosions alone.

From the point of view of SN Ia nucleosynthesis, the main difference between near-\Mch\ and sub-\Mch\ 
explosions is whether the WD core is dense enough for electron capture ($p + e^- \rightarrow n + \nu_e$) 
to take place during the initial phase of the explosion. The threshold density for this 
reaction ($\rho_c$ $\approx$ $10^8$\,g\,cm$^{-3}$) is only achieved when the WD mass is close to \Mch.
A distinguishing characteristic of the two models is, therefore, the production efficiency of neutron-rich species, 
like Ni and Mn, that is higher in the near-\Mch\ scenario\cite{Maeda10a,Seitenzahl13b,Yamaguchi15}. 
We may exploit this distinction to identify the dominant type of SN Ia progenitors in galaxy clusters by measuring 
the abundance of the Fe-peak elements in the ICM. The results may apply globally, since rich galaxy clusters 
represent a scale sufficiently massive to be representative of the universe as a whole.

Here we model the SXS spectrum of the Perseus Cluster in the 1.8--9.0 keV band (Fig.\,1a) 
with an optically thin thermal plasma in collisional ionization equilibrium using the latest atomic codes 
(AtomDB v.3.0.8 and SPEX v.3.03). 
The emission from the active galactic nucleus (AGN) of the cD galaxy NGC\,1275 is taken 
into account by adding a power law and redshifted Fe~{\footnotesize I} K$\alpha_1$ and K$\alpha_2$ lines. 
Details about the analysis and systematic uncertainty assessment are described in the Methods section. 
Our constraints on the elemental abundances, with respect to Fe, are shown in Fig.\,2 (red circles). 
These are fully consistent with the solar abundance ratios\cite{Lodders09}.

Fig.\,2 also plots previously measured ICM abundances of the Perseus Cluster core as well as the average 
of 44 objects including galaxy clusters, groups, and elliptical galaxies from XMM-Newton observations 
(blue triangles and squares)\cite{Mernier16a}. 
This plot highlights some important differences between the measurements. 
First, the SXS-measured abundances have statistical uncertainties comparable to the XMM-Newton results 
from the combined data of the 44 objects, despite a 15-times shorter exposure and a much smaller field of view. 
Second, while the abundances of Si, S, Ar, and Ca are identical between the two studies, 
the earlier measurements systematically obtained supersolar abundances of the Fe-peak element 
from both the Perseus Cluster and the 44-object average.

Previous X-ray studies of clusters and elliptical galaxies often obtained a supersolar Ni/Fe ratio, 
leading the study authors to argue for differences in the nature of SNe Ia between the early-type 
galaxies and the Milky Way\cite{Dupke00,Gastaldello04,dePlaa07,DeGrandi09,Mernier16b}. 
By contrast, optical spectra of old stars in early-type galaxies indicate that the relative abundances among 
the Fe-peak elements are consistent with the solar value (see yellow stars in Fig.\,2)\cite{Conroy14}. 
Our new X-ray measurement relieves this discrepancy and strongly suggests that the average nature of 
SNe Ia is independent of the star formation history of their host galaxies. 
This robust result, unaffected by complicated radiative transfer that may lend uncertainty to optical studies, 
is obtained by an accurate determination of the Ni abundance primarily based on the intensity of its resonance
emission line that is easily resolved from the Fe He$\beta$ line and other weak emission of Fe XXIV and Fe XXV.

Since Cr and Mn abundances of individual objects were not constrained by the previous XMM-Newton 
observations\cite{Mernier16a}, we cannot exclude the possibility that sample variance leads at least in part 
to the discrepancy between the two studies.
Nevertheless, we demonstrate in Methods that high resolution spectroscopy is 
essential for robust measurements of these abundances. In short, only the SXS can clearly separate 
the weak resonance lines from the continuum component, enabling abundance measurements 
much less subject to systematic uncertainties in spectral modeling.
The high-resolution SXS data have also stimulated the development of atomic models, 
reducing the uncertainties in the modeled line emissivities and improving 
the accuracy of the abundances with respect to the previous work.

Fig.\,3 compares the SXS-measured abundances of the Fe-peak elements (black data points) with 
theoretical yields from the latest three-dimensional calculations of the near-\Mch\ SN Ia\cite{Seitenzahl13a} 
(blue region) and sub-\Mch\ merger\cite{Pakmor12} (green region). 
We also consider a one-dimensional explosion of a single 1.0\,\Msun\ WD\cite{Woosley11}
(gray region) as an alternative example of a sub-\Mch\ SN Ia model.
All of these models predict typical SN Ia brightness and a synthesized $^{56}$Ni mass of $\sim$\,0.6\,\Msun. 
In addition, contributions of core-collapse (CC) SNe are accounted for in each model given in the figure, 
utilizing mass-dependent yields\cite{Nomoto13} averaged over the Salpeter initial mass function (IMF). 
We allow a conservatively wide range for the CC SN fraction, 
$f_{\rm CC}$ $\equiv$ $N_{\rm CC}/(N_{\rm Ia}+N_{\rm CC})$ = 0.6--0.9 
(typical for cluster cores\cite{Sato07,dePlaa07,Bulbul12,Matsushita13}), 
instead of constraining an actual value from our observation (see Methods for more details). As expected, 
the near-\Mch\ model predicts higher abundances of Mn and Ni owing to the efficient electron capture. 
The observed abundance pattern disfavors a hypothesis that all SNe Ia involve sub-\Mch\ WD, 
and prefers the combination of the near-\Mch\ and sub-\Mch\ SNe Ia with roughly equal numbers 
(red region in the figure). We also find that our result starkly contrasts with previous
claims\cite{Dupke00,Mernier16b}, where introduction of rather non-standard full-deflagration SN Ia models 
was required to understand a Ni/Fe ratio that was estimated to be much higher than our measurement. 
In Methods, we investigate other current SN Ia and CC models and find that our main conclusion remains valid, 
although an exact ratio of near-\Mch\ to sub-\Mch\ contributions may depend on the model details.

The Hitomi SXS observation has demonstrated the power of high-resolution X-ray spectroscopy: through measurement 
of the chemical enrichment of a single object, new insight has been gained into fundamental phenomena 
shaping the present-day universe. A common abundance pattern between the solar neighborhood and 
the Perseus Cluster suggests that the Sun's chemical composition is likely to be a good indicator of 
the average SN Ia nature in the universe.
It is extremely important to scrutinize other environments like outskirts of galaxy clusters\cite{Werner13} 
at high spectral resolution, a task left for future X-ray observatories.

\clearpage


\section*{References}

\clearpage

\begin{addendum}


\item[Acknowledgements] 
Acknowledgements are provided in the Supplementary Information.


\item[Author Contributions] 
H. Yamaguchi wrote the manuscript. 
H. Yamaguchi, S. Nakashima, A. Simionescu, E. Bulbul, and M. Loewenstein analyzed the data specifically for this project. 
H. Yamaguchi, K. Matsushita, M. Loewensterin, A. Simionescu, S. Nakashima, K. Sato, and R. Mushotzky discussed the results. 
Y. Ishisaki confirmed the reliability of the observed results based on his expertise in the SXS signal processing system. The science goals of Hitomi were discussed and developed over more than 10 years by the ASTRO-H Science Working Group (SWG), all members of which are authors of this manuscript. All the instruments were prepared by joint efforts of the team. Calibration of the Perseus dataset was carried out by members of the SXS team. The manuscript was subject to an internal collaboration-wide review process. All authors reviewed and approved the final version of the manuscript.


\item[Author Information] 
Reprints and permissions information is available at www.nature.com/reprints.
The authors declare no competing financial interests.
Correspondence and requests for materials should be addressed to 
H.\ Yamaguchi (hiroya.yamaguchi@nasa.gov) 
and K.\ Matsushita (matusita@rs.kagu.tus.ac.jp).


 \item[Hitomi Collaboration]
Felix Aharonian$^{1,2,3}$,
Hiroki Akamatsu$^{4}$,
Fumie Akimoto$^{5}$,
Steven W. Allen$^{6,7,8}$,
Lorella Angelini$^{9}$,
Marc Audard$^{10}$,
Hisamitsu Awaki$^{11}$,
Magnus Axelsson$^{12}$,
Aya Bamba$^{13,14}$,
Marshall W. Bautz$^{15}$
% ...

\end{addendum}


\begin{affiliations}
	\item Dublin Institute for Advanced Studies, 31 Fitzwilliam Place, Dublin 2, Ireland
        \item Max-Planck-Institut f\"ur Kernphysik, P.O. Box 103980, 69029 Heidelberg, Germany
        \item Gran Sasso Science Institute, viale Francesco Crispi, 7 67100 L'quila (AQ), Italy
	\item SRON Netherlands Institute for Space Research, Sorbonnelaan 2, 3584 CA Utrecht, The Netherlands
        \item Institute for Space-Earth Environmental Research, Nagoya University, Furo-cho, Chikusa-ku, Nagoya, Aichi 464-8601, Japan
	\item Kavli Institute for Particle Astrophysics and Cosmology, Stanford University, 452 Lomita Mall, Stanford, CA 94305, USA
	\item Department of Physics, Stanford University, 382 Via Pueblo Mall, Stanford, CA 94305, USA
	\item SLAC National Accelerator Laboratory, 2575 Sand Hill Road, Menlo Park, CA 94025, USA
	\item NASA, Goddard Space Flight Center, 8800 Greenbelt Road, Greenbelt, MD 20771, USA
	\item Department of Astronomy, University of Geneva, ch. d'\'Ecogia 16, CH-1290 Versoix, Switzerland
	\item Department of Physics, Ehime University, Bunkyo-cho, Matsuyama, Ehime 790-8577, Japan
	\item Department of Physics and Oskar Klein Center, Stockholm University, 106 91 Stockholm, Sweden
	\item Department of Physics, The University of Tokyo, 7-3-1 Hongo, Bunkyo-ku, Tokyo 113-0033, Japan
	\item Research Center for the Early Universe, School of Science, The University of Tokyo, 7-3-1 Hongo, Bunkyo-ku, Tokyo 113-0033, Japan
	\item Kavli Institute for Astrophysics and Space Research, Massachusetts Institute of Technology, 77 Massachusetts Avenue, Cambridge, MA 02139, USA
        \item Smithsonian Astrophysical Observatory, 60 Garden St., MS-4. Cambridge, MA 02138, USA
	\item Lawrence Livermore National Laboratory, 7000 East Avenue, Livermore, CA 94550, USA
	\item Department of Physics and Astronomy, Wayne State University,  666 W. Hancock St, Detroit, MI 48201, USA
	\item Astronomy Department, Yale University, New Haven, CT 06520-8101, USA
	\item Physics Department, Yale University, New Haven, CT 06520-8120, USA
	\item Centre for Extragalactic Astronomy, Department of Physics, University of Durham, South Road, Durham, DH1 3LE, UK
	\item Japan Aerospace Exploration Agency, Institute of Space and Astronautical Science, 3-1-1 Yoshino-dai, Chuo-ku, Sagamihara, Kanagawa 252-5210, Japan
	\item Department of Astronomy, Kyoto University, Kitashirakawa-Oiwake-cho, Sakyo-ku, Kyoto 606-8502, Japan
	\item The Hakubi Center for Advanced Research, Kyoto University, Kyoto 606-8302, Japan
	\item Department of Physics, Tokyo Metropolitan University, 1-1 Minami-Osawa, Hachioji, Tokyo 192-0397, Japan
	\item Institute of Astronomy, University of Cambridge, Madingley Road, Cambridge, CB3 0HA, UK
	\item Faculty of Mathematics and Physics, Kanazawa University, Kakuma-machi, Kanazawa, Ishikawa 920-1192, Japan
	\item School of Science, Hiroshima University, 1-3-1 Kagamiyama, Higashi-Hiroshima 739-8526, Japan
        \item Fujita Health University, Toyoake, Aichi 470-1192, Japan
	\item Physics Department, University of Miami, 1320 Campo Sano Dr., Coral Gables, FL 33146, USA
	\item Department of Astronomy and Physics, Saint Mary's University, 923 Robie Street, Halifax, NS, B3H 3C3, Canada
	\item Department of Physics and Astronomy, University of Southampton, Highfield, Southampton, SO17 1BJ, UK
	\item Laboratoire APC, 10 rue Alice Domon et L\'eonie Duquet, 75013 Paris, France
	\item CEA Saclay, 91191 Gif sur Yvette, France
	\item European Space Research and Technology Center, Keplerlaan 1 2201 AZ Noordwijk, The Netherlands
        \item Department of Physics and Astronomy, Aichi University of Education, Aichi 448-8543, Japan
	\item Department of Physics, Tokyo University of Science, 2641 Yamazaki, Noda, Chiba, 278-8510, Japan
	\item Department of Physics, University of Maryland Baltimore County, 1000 Hilltop Circle, Baltimore,  MD 21250, USA
	\item Department of Applied Physics and Electronic Engineering, University of Miyazaki, 1-1 Gakuen Kibanadai-Nishi, Miyazaki, 889-2192, Japan
	\item Department of Physics, Nagoya University, Furo-cho, Chikusa-ku, Nagoya, Aichi 464-8602, Japan
	\item Department of Earth and Space Science, Osaka University, 1-1 Machikaneyama-cho, Toyonaka, Osaka 560-0043, Japan
	\item Department of Physics, Kwansei Gakuin University, 2-1 Gakuen, Sanda, Hyogo 669-1337, Japan
	\item Department of Physics, Rikkyo University, 3-34-1 Nishi-Ikebukuro, Toshima-ku, Tokyo 171-8501, Japan
	\item Department of Physics and Astronomy, Rutgers University, 136 Frelinghuysen Road, Piscataway, NJ 08854, USA
	\item Meisei University, 2-1-1 Hodokubo, Hino, Tokyo 191-8506, Japan
        \item Leiden Observatory, Leiden University, PO Box 9513, 2300 RA Leiden, the Netherlands
	\item Research Institute for Science and Engineering, Waseda University, 3-4-1 Ohkubo, Shinjuku, Tokyo, 169-8555, Japan
	\item Department of Physics, Chuo University, 1-13-27 Kasuga, Bunkyo, Tokyo 112-8551, Japan
	\item Department of Physics, Tokyo Institute of Technology, 2-12-1 Ookayama, Meguro-ku, Tokyo 152-8550, Japan
	\item Department of Physics, Toho University,  2-2-1 Miyama, Funabashi, Chiba 274-8510, Japan
	\item Department of Physics, Kyoto University, Kitashirakawa-Oiwake-Cho, Sakyo, Kyoto 606-8502, Japan
	\item European Space Astronomy Center, Camino Bajo del Castillo, s/n.,  28692 Villanueva de la Ca{\~n}ada, Madrid, Spain
	\item Universities Space Research Association, 7178 Columbia Gateway Drive, Columbia, MD 21046, USA
	\item National Science Foundation, 4201 Wilson Blvd, Arlington, VA 22230, USA
	\item Department of Electronic Information Systems, Shibaura Institute of Technology, 307 Fukasaku, Minuma-ku, Saitama-shi, Saitama 337-8570, Japan
	\item Department of Astronomy, University of Maryland, College Park, MD 20742, USA
	\item Space Telescope Science Institute, 3700 San Martin Drive, Baltimore, MD 21218, USA
        \item Institute of Physical and Chemical Research, 2-1 Hirosawa, Wako, Saitama 351-0198
	\item Department of Physics, Tokyo University of Science, 1-3 Kagurazaka, Shinjuku-ku, Tokyo 162-8601, Japan
        \item Department of Physics, University of Wisconsin, Madison, WI 53706, USA
	\item Department of Physics and Astronomy, University of Waterloo, 200 University Avenue West, Waterloo, Ontario, N2L 3G1, Canada
	\item Department of Astronomy, University of Michigan, 1085 South University Avenue, Ann Arbor, MI 48109, USA
	\item Okinawa Institute of Science and Technology Graduate University, 1919-1 Tancha, Onna-son Okinawa, 904-0495, Japan
        \item Hiroshima Astrophysical Science Center, Hiroshima University, Higashi-Hiroshima, Hiroshima 739-8526, Japan
	\item Faculty of Liberal Arts, Tohoku Gakuin University, 2-1-1 Tenjinzawa, Izumi-ku, Sendai, Miyagi 981-3193
	\item Faculty of Science, Yamagata University, 1-4-12 Kojirakawa-machi, Yamagata, Yamagata 990-8560, Japan
	\item Department of Physics, Nara Women's University, Kitauoyanishi-machi, Nara, Nara 630-8506, Japan
	\item Department of Teacher Training and School Education, Nara University of Education, Takabatake-cho, Nara, Nara 630-8528, Japan
	\item Frontier Research Institute for Interdisciplinary Sciences, Tohoku University,  6-3 Aramakiazaaoba, Aoba-ku, Sendai, Miyagi 980-8578, Japan
	\item Astronomical Institute, Tohoku University, 6-3 Aramakiazaaoba, Aoba-ku, Sendai, Miyagi 980-8578, Japan
	\item Astrophysics Laboratory, Columbia University, 550 West 120th Street, New York, NY 10027, USA
	\item Department of Physics and Astronomy, University of Manitoba, Winnipeg, MB R3T 2N2, Canada
	\item Department of Physics and Mathematics, Aoyama Gakuin University, 5-10-1 Fuchinobe, Chuo-ku, Sagamihara, Kanagawa 252-5258, Japan
	\item Astronomical Observatory of Jagiellonian University, ul. Orla 171, 30-244 Krak\'ow, Poland
	\item RIKEN Nishina Center, 2-1 Hirosawa, Wako, Saitama 351-0198, Japan
        \item Max Planck Institute for extraterrestrial Physics, Giessenbachstrasse 1, 85748 Garching, Germany
	\item Department of Physics, Saitama University, 255 Shimo-Okubo, Sakura-ku, Saitama, 338-8570, Japan
        \item Department of Physics, University of Rome ``Tor Vergata'', Via della Ricerca Scientifica 1, I-00133 Rome, Italy
	\item Faculty of Education, Shizuoka University, 836 Ohya, Suruga-ku, Shizuok
a 422-8529, Japan
	\item Faculty of Health Sciences, Nihon Fukushi University , 26-2 Higashi Haemi-cho, Handa, Aichi 475-0012, Japan
	\item MTA-E\"otv\"os University Lend\"ulet Hot Universe Research Group, P\'azm\'any P\'eter s\'et\'any 1/A, Budapest, 1117, Hungary
	\item Department of Theoretical Physics and Astrophysics, Faculty of Science, Masaryk University, Kotl\'a\v{r}sk\'a 2, Brno, 611 37, Czech Republic
	\item Department of Physics and Astronomy, University of Utah, 115 South 1400 East, Salt Lake City, Utah 84112, USA
	\item The Johns Hopkins University, Homewood Campus, Baltimore, MD 21218, USA

\end{affiliations}


\clearpage


\begin{figure}
  \begin{center}
	\includegraphics[width=16cm]{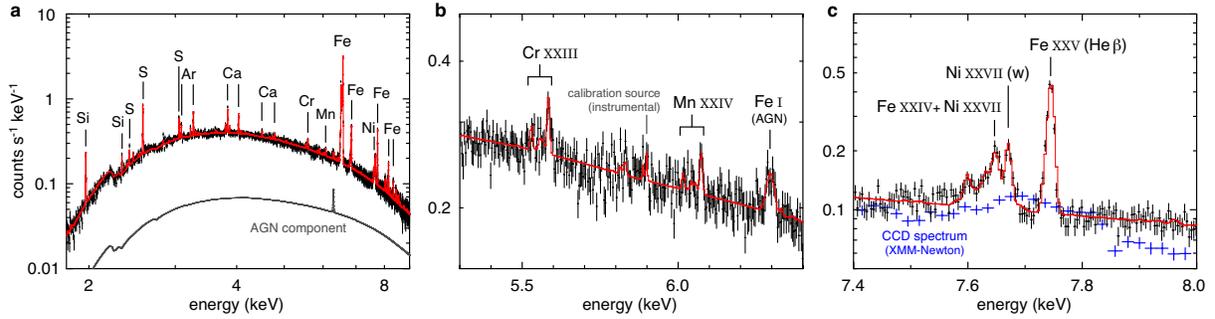}	
\caption{
{\bf The Hitomi/SXS spectra of the Perseus Cluster.}  
 (a) The spectrum (black) in the 1.8--9.0\,keV band modeled with an optically thin thermal plasma based on 
  the atomic code AtomDB (red). 
  The error bars are at a 1$\sigma$ confidence level. 
  The emission from NGC\,1275 (AGN) is indicated by the gray curve. 
  The spectrum is rebinned by 4\,eV for clarity, though 1-eV bins were used for fitting.  \ 
  (b) The zoom-in spectrum in the 5.3--6.4\,keV band, where the emission from He-like Cr and Mn are detected. 
  The red-shifted Fe~{\footnotesize I} fluorescence from the AGN is resolved as well. \
  (c) The same in the 7.4--8.0 keV band, highlighting the Ni~{\footnotesize XXVII} resonance ({\it w}) 
  line clearly separated from the stronger Fe~{\footnotesize XXV} He$\beta$ and other emission. 
  This enables the first accurate measurement of the Ni abundance in a galaxy cluster. 
   For comparison, an XMM-Newton spectrum extracted from the same spatial region is shown as 
  the blue data points. \label{full}}
  \end{center}
\end{figure}

\clearpage

\begin{figure}
  \begin{center}
	\includegraphics[width=8.5cm]{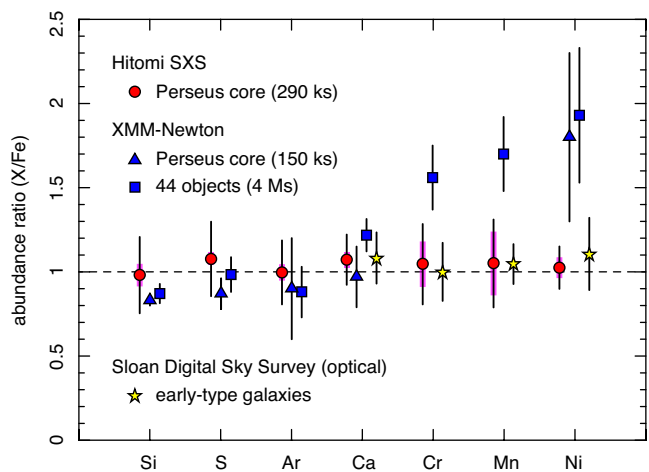}	
\caption{
{\bf Elemental abundances of the Perseus Cluster.}  
The values are relative to the solar abundances\cite{Lodders09} with respect to Fe. 
The red circles represent the SXS measurements with error bars of typical statistical uncertainty
at a 1$\sigma$ confidence level (thick magenta) and systematic uncertainty due to the model selection 
(thin black: see Methods for details). 
The blue triangles and squares represent the XMM-Newton results from the Perseus Cluster core 
and the integrated data of 44 objects, respectively\cite{Mernier16a}. 
The yellow stars show the optical measurements of stellar abundances
in early-type galaxies from the Sloan Digital Sky Survey\cite{Conroy14},  
where velocity dispersion dependence and systematic errors of 0.05 dex are
taken into account in the error bars.  Si is not shown because its
abundance is highly sensitive to the velocity dispersion. S and Ar
abundances are unavailable in the optical study. 
\label{abund}}
  \end{center}
\end{figure}

\clearpage

\begin{figure}
  \begin{center}
	\includegraphics[width=8.5cm]{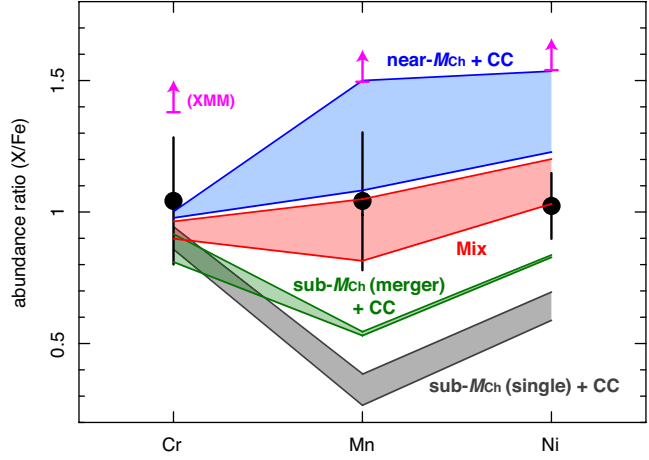}
\caption{
{\bf Comparison between the observed abundances and theoretical calculations
for the Fe-peak elements.} 
The black data points are identical to the red circles in Fig.\,2, the SXS-measured 
abundance ratios relative to the solar abundances\cite{Lodders09}. 
The error bars include both statistical uncertainty at a 1$\sigma$ confidence level 
and systematic uncertainty. The magenta arrows indicate the 1$\sigma$ lower limit of 
the XMM-Newton measurements for the 44 objects\cite{Mernier16a}. 
The blue, green, and gray regions represent the theoretical predictions for SNe Ia from 
the near-\Mch\ delayed-detonation explosion\cite{Seitenzahl13a}, 
sub-\Mch\ violent merger\cite{Pakmor12}, 
and single sub-\Mch\ WD\cite{Woosley11}, respectively. 
In each model, contributions from CC SNe\cite{Nomoto13} are also taken into account (see text). 
The red region assumes equal contributions of the near-\Mch\ SNe Ia and sub-\Mch\ violent mergers, 
providing a reasonable fit to the data (although the exact ratio between the two SN Ia types is 
subject to some uncertainties in the model details).
\label{fepeak}}
  \end{center}
\end{figure}

\clearpage


\begin{methods}

\subsection{Observations and Data Reduction:}

The Hitomi observations of the Perseus Cluster core were performed using the SXS in the sequences 
summarized in Extended Data Table 1. The SXS field of view (FoV) of each sequence is indicated in Extended Data Figure 1. 
The data from the first four sequences, whose aim points were almost identical, 
were used in our previous work as well\cite{Hitomi16}. The spacecraft attitude was slightly different 
for the last sequence, so that the nucleus of NGC\,1275 was observed using the central pixels of 
the SXS. The aperture window consisting of a 262-$\mu$m Be filter and several contaminant materials 
\cite{Eckart16} was not opened before the loss of the mission. 
This filter significantly attenuated the SXS effective area especially in the soft X-ray band, 
limiting the SXS bandpass to above $\sim$\,1.8\,keV.

The data reduction was made with public tools provided by NASA's HEASARC. 
We used cleaned event data of the latest release version with the standard screening 
for the post-pipeline processes\cite{Angelini16}. The spectral analysis was performed 
using only GRADE Hp (high-resolution primary) events that have the best energy resolution. 
The redistribution matrix file (RMF) was generated with the extra-large size option, which accounts for 
all components of the line spread function, including the main peak, low-energy exponential tail, escape peaks, 
and electron-loss continuum\cite{Eckart16,Leutenegger16}. The full width at half maximum (FWHM) of 
the main-peak component was measured to be 4.9\,eV for the $^{55}$Fe calibration source\cite{Kelley16,Porter16}.

\subsection{Additional Gain Correction:}

Because of the short life of the mission, opportunities for onboard calibration were limited. This caused 
some uncertainty in the detector gain (pulse height--energy conversion factors), particularly at the energies 
far from the Mn \Ka\ calibration lines at 5.9\,keV (in a calibration pixel irradiated by a $^{55}$Fe source). 
We thus applied the following gain calibration and correction using the Perseus data themselves.
 
First, we model the Fe He$\alpha$ complex with an ionization equilibrium plasma 
for each pixel in each sequence (combining the second through fourth sequences in Extended Data Table 1, 
since these were parts of a continuous observation with almost identical instrumental conditions), 
and scaled the spectrum with a linear function such that the Fe He$\alpha$ energies match 
the theoretical values at the redshift of NGC\,1275 ($z = 0.01756$)\cite{Ferruit97}. 
We then merged the data of all the pixels and measured the X-ray energies of detected lines. 
The differences between the measured and theoretical energies are plotted in Extended Data Figure 2. 
The discrepancy, while small, increases toward lower and higher energies with respect to 
the calibration source line (i.e., 5.9\,keV). We empirically fit these plots with a parabolic function, 
and then updated the pulse invariant spectral channel of each detected event using the derived 
coefficients. Readers are cautioned that this empirical correction should not be used outside of 
the range of the fit; in particular, the actual gain error must be almost zero at the energies near 0\,eV.
The data from all sequences were then combined to increase the photon statistics. 
Hereafter, we use this merged, gain-corrected spectrum. 
We also appropriately took into account the off-axis effective area of the Soft X-ray Telescope (SXT) 
\cite{Okajima16}, when generating the telescope response for the merged data.

\subsection{Spectral Analysis:}

We analyzed the SXS spectrum in the 1.8--9.0\,keV band with an energy bin size of 1\,eV. 
The spectral fitting was made using the C statistic\cite{Cash79} without subtracting any background 
component, since its level is negligibly low ($\sim$\,$7 \times 10^{-4}$\,counts~s$^{-1}$\,keV$^{-1}$ for 
the entire field of view), with even its strongest emission lines well below the source flux 
in the 1.8--9.0 keV energy band. 
In fact, no significant change in the spectral parameters is found, if we 
fit the source spectrum by simultaneously modeling the instrumental 
background data extracted from the night-Earth observations. 
The cosmic X-ray background is also negligible at this cluster core region; 
well below 1\% of the source emission over the entire energy band\cite{Tamura09}.

We fit the spectrum of the Perseus Cluster with a single-temperature optically thin thermal plasma model 
({\tt bvvapec} model in the XSPEC package) based on the latest version of the atomic database, 
AtomDB v.3.0.8\cite{Foster12}. The fitted parameters included the electron temperature ($kT_e$), redshift ($z$), 
turbulence velocity ($v_t$), emission measure, and the elemental abundances of Si, S, Ar, Ca, Cr, Mn, Fe, 
and Ni relative to the solar values (Extended Data Table 2)\cite{Lodders09}. We included a power-law component 
and redshifted lines of Fe~{\footnotesize I} K$\alpha$ fluorescence (6.4\,keV at the rest frame)
to account for the emission from the AGN of NGC\,1275\cite{Churazov04}. 
The photon index and flux of the power law component were determined to be $\Gamma \approx 1.9$ 
and $F_{2-10\,{\rm keV}} \approx 3 \times 10^{-11}$\,erg~s$^{-1}$\,cm$^{-2}$ using an AGN-dominated 
spectrum derived by SXS image analysis decomposing AGN and ICM emissions, 
and fixed to these values in the analysis of the ICM spectrum (Fig.\,1) that was extracted from the entire SXS array.
A foreground absorption column (\NH) was fixed at $1.38 \times 10^{21}$\,cm$^{-2}$ \cite{Kalberla05}. 
The possible effect of resonance scattering (RS)\cite{Zhuravleva13,Hitomi16} was accounted by 
adding a Gaussian at the energy of the Fe~{\footnotesize XXV} resonance line with a negative flux. 
Weak $^{55}$Fe calibration source leakage events were taken into account by adding narrow Gaussians at 
the theoretical energies of the Mn K$\alpha$ lines, although this has no impact on our analysis results. 
With this model (hereafter ``Model A1''), we obtained best-fit values of $kT_e$ = 3.97 $\pm$ 0.02~keV 
and the absolute Fe abundance (i.e., the Fe/H number ratio relative to the solar values)  
of 0.63 $\pm$ 0.01~solar, with a C-statistic and $\chi ^2$ of 7483 and 7862, respectively 
(7180 degrees of freedom). 
The relative abundances of the other elements (with respect to Fe) are shown in Extended Data Figure 3. 
Note that the uncertainty in our gain correction is less than 1\,eV at energies near the Mn \Ka\ calibration lines 
(Extended Data Figure 2), and thus its effect is negligible for the determination of the Fe-peak element abundances.

We carefully estimated systematic uncertainties in the measured abundances by introducing different 
models and assumptions. First, we excluded the RS correction, i.e., the negative-flux Fe~{\footnotesize XXV} 
line (Model A2).  
This did not substantially change the relative abundances, confirming suggestions in previous work 
on this object\cite{Churazov04,Tamura09}. We also fit the spectrum with two-temperature models, 
with and without the RS effect (Models A3 and A4, respectively). In these models, the parameters 
other than the temperatures and emission measures were linked between the two components. 
We obtained best-fit temperatures of $kT_{e1}$ = 4.04 $\pm$ 0.05~keV and $kT_{e2}$ = 1.60 $\pm$ 0.27~keV 
with 2--10-keV flux ratio ($F_1/F_2$) of 33.5 for Model A3, and similar values for Model A4. 
This indicates that the 4-keV component dominates over the entire SXS band and that the one-temperature 
modeling is already a good approximation for the observed region in this bandpass, although the presence 
of a multi-temperature plasma was previously inferred for this cluster\cite{Sanders07,Zhuravleva13}. 
already a good approximation for the observed region in this bandpass, although the presence of a 
multi-temperature plasma was previously inferred for this cluster\cite{Sanders07,Zhuravleva13}. 
We also treated the absorption columns and the AGN spectral index and flux as free parameters, 
and confirmed no significant change in the relative abundances among the Fe-peak elements. 
Finally, we used the SPEX atomic code v.3.03\cite{Kaastra96} to fit the same spectral data with the same 
model components and assumptions (Models S1--S4, equivalent to Models A1--A4, respectively). 
The measured abundance ratios for each model are summarized in Extended Data Figure 3. 
The ranges between the minimum and maximum values among Models A1--A4 and S1--S4 are 
given in Fig.\,2 as the uncertainty for the abundance of each element. 
The systematic uncertainties owing to the different atomic codes and assumptions are 
larger than the statistical errors but reasonably small for most of the elements.
All the metal abundances are found to be fairly consistent with the solar values\cite{Lodders09}. 
There are no significant differences in abundances derived from analysis of a region excluding 
the $2' \times 2'$ box centered on the AGN of NGC\,1275.

We have found that the abundance ratios of Cr/Fe, Mn/Fe, and Ni/Fe 
are systematically lower than those determined in recent XMM-Newton studies\cite{Mernier16a}. 
Because an old plasma model (SPEX v.2.05) was used in this previous work, we also fit the SXS 
spectrum using that model for direct comparison. The results from one- and two-temperature modeling 
with the RS correction are given in Extended Data Figure 3 (Models S$'$1 and S$'$3, respectively) 
and Extended Data Figure 4 (red diamonds) with the combined uncertainty ranges. 
Cr and Mn abundances are not presented, because the SPEX v.2.05 atomic code does not contain 
emission from these elements --- in the previous work, abundances of these elements 
were calculated by referring to emissivity data in an early development version of SPEX v.3.
The Ni abundance determined from this old atomic model is slightly higher than from the latest one 
(SPEX v.3.03), but still lower than the XMM-Newton results. 
In fact, there is little difference in the Ni-He$\alpha$ emissivity itself between SPEX v.2.05 and v.3.03. 
We find significant differences between the two SPEX versions in the line 
emissivities of Fe XXIV and Fe XXV complex at the rest frame energies of 7.6--7.9\,keV.
Given that these emission cannot be separated from the Ni resonance line in CCD spectra, 
the Ni abundance might have been biased in the previous measurements.

Since Cr and Mn are rarely detected from individual objects with CCD observations, it is not obvious 
whether the supersolar abundances derived from the integrated XMM-Newton data of the 44 objects 
are real or biased. On the other hand, Suzaku observations (with similar CCDs) detected these elements from 
the same Perseus core region as in this work\cite{Tamura09}. The Suzaku-measured abundances, 
converted to the same scale using the up-to-date solar abundance table\cite{Lodders09}, 
are compared with the Hitomi and XMM-Newton results in Extended Data Figure 4 (green squares). 
This earlier measurement of the Mn/Fe ratio is significantly lower than ours, further motivating 
the following demonstration of the robustness of our measurements compared to that of CCD observations.

Extended Data Figure 5(a) shows the SXS spectrum near the Cr and Mn emission lines, 
of which equivalent widths are only a few electron volts. 
The red line indicates our best-fit model (Model A1) but with Cr and Mn abundances set to zero. 
As shown in the bottom panel of the figure, the photon count ratios between the line peak and 
the local continuum level is $\sim$\,1.2 for these weak emission lines in this high-resolution spectrum. 
Extended Data Figure 5(b) is a similar plot but the spectrum is convolved to the resolution of CCDs 
using a representative XMM-Newton response function. Unlike the SXS spectrum, the peak-to-continuum 
level ratios for the Cr and Mn emission are extremely low (only a few percent above unity). Moreover, 
the emission lines no longer have a sharp profile, implying the difficulty in separating lines from continuum. 
In fact, if we fit this simulated CCD spectrum with a model with 1\% higher/lower continuum normalization, 
the line components with their broad profiles `compensate' for the excess/lack of continuum flux by 
requiring $\sim$\,50\% lower/higher values of the Cr/Fe and Mn/Fe abundance ratios. 
The high resolution SXS spectrum is much less subject to such systematic uncertainties, 
since the line and continuum intensities are measured almost independently and hence a slight 
over- or under-estimation of the continuum level has little effect on the abundance measurement.
This point is more quantitatively illustrated in Extended Data Figure 6, the result of our test analysis.

\subsection{Comparison with SN Nucleosynthesis Models:}

The measured abundances of the Fe-peak elements are compared with
theoretical predictions to address the nature of SNe Ia that likely
contributed to the chemical enrichment in the Perseus Cluster.  As
prototype SN Ia models, we select the latest three-dimensional
calculations ``N100''\cite{Seitenzahl13a} and
``1.1\_0.9''\cite{Pakmor12}.  The former assumes a delayed-detonation
explosion of a near-\Mch\ WD with 100 deflagration ignition sites. The
latter assumes the violent merger of two sub-\Mch WDs with masses of
1.1\Msun\ and 0.9\Msun\, and subsequent full detonation of the primary
(more massive) WD.  Both models successfully replicate typical
observables of SNe Ia, including the average maximum brightness and
synthesized $^{56}$Ni mass of $\sim$\,0.6\,\Msun.  The pre-explosion
WD is composed of 47.5\% $^{12}$C, 50\% $^{16}$O, and 2.5\% $^{22}$Ne
by mass, which corresponds to nearly solar metallicity for the progenitor.  
As another example of a sub-\Mch\ explosion, we choose the ``10HC''
model\cite{Woosley11}, which assumes an explosion of a single C--O WD
with a mass of 1.0\,\Msun\ accreting helium at a rate $\dot{M} = 4.0
\times 10^{-8}$\,\Msun\,yr$^{-1}$. An initial detonation ignited at
the helium layer triggers a second detonation in the CO core,
resulting in a complete explosion of the WD with a kinetic energy of
$1.2 \times 10^{51}$\,erg and $^{56}$Ni mass of $\sim$\,0.64\,\Msun,
as typically inferred for SNe Ia.

To account for the CC SN contributions, we consider mass-dependent
yields\cite{Nomoto13} weighted by the Salpeter IMF ($\alpha$ = 2.35),
with the assumption that 50\% of $\geq$\,25\,\Msun\ massive stars
explode as hypernovae. Since SNe Ia efficiently produce Fe,
whereas SNe CC dominate $\alpha$-element production, the SXS spectra
we extracted might be used to constrain the SN Ia/CC ratio in the
Perseus Cluster. However, we instead allow a conservatively wide range
for the CC SN fraction, $f_{\rm CC}$ $\equiv$ $N_{\rm CC}/(N_{\rm
Ia}+N_{\rm CC})$ =
0.6--0.9\cite{Sato07,dePlaa07,Bulbul12,Matsushita13,Finoguenov02,Simionescu09,Ezer17},
rather than determining the actual $f_{\rm CC}$ value. This choice was
made because (1) the lighter elements that are most sensitive to
$f_{\rm CC}$ (i.e., O, Ne, Mg) were not detected due to the
attenuation of soft X-rays by the closed aperture window; (2) the
measured abundances of the intermediate $\alpha$-burning elements,
unlike those of the Fe-peak elements, are dominated by systematic,
rather than the statistical, uncertainties (Extended Data Figure 3);
and (3) the primary origins of Ar and Ca is currently under
debate\cite{Mulchaey14,Mernier16b}.  Future high-resolution X-ray
spectroscopy with sensitivity to softer X-rays will improve the
accuracy of the abundances of the lighter elements, as well as of the
ICM spectral model, hence enabling better constrains on the SN Ia/CC
ratio.  We emphasize that, in contrast to the intermediate
$\alpha$-burning elements, the abundances of the Fe-peak elements are
robustly determined with little model dependency (Extended Data Figure
3). As a result, the main conclusions of this paper are not affected
by any of the issues described above.

The abundance ratios predicted by the model calculations are given in Fig.\,3. 
Because of the efficient electron capture as well as the low entropy
freeze-out from nuclear statistical equilibrium\cite{Seitenzahl13b},
higher abundances of Mn and Ni are expected in the near-\Mch\ SNe Ia.
We also test other combinations of SN models as well as different IMF
slopes (for CC SNe).  Extended Data Table 3 summarizes the mass ratios among the
Fe-peak elements and Fe yields (in \Msun) predicted by the various SN
Ia models we investigated\cite{Seitenzahl13a,Fink14,Maeda10a,Travaglio04,Travaglio11,Pakmor12,Woosley11,Bravo12,Yamaguchi15,Tsujimoto12}. 
Since this paper exclusively discusses the products of electron capture, we
consider only recent calculations that were based on up-to-date weak
interaction rates\cite{Langanke00}.  For CC SN models, we use
different IMF slopes ($\alpha$ = 2.0 and 2.7) and assume that all
10--50\,\Msun\ stars explode as normal SNe without any hypernova
contribution.  These results are summarized in Extended Data Table 4.  We reach
essentially the same conclusion described in the main text, i.e.,
higher mass ratios of Mn/Fe and Ni/Fe are always expected from
near-\Mch\ SNe Ia (Extended Data Table 3), and a combination of near-\Mch\ and
sub-\Mch\ SNe Ia naturally explains the observed abundance pattern of
the Fe-peak elements independently of contributions from CC SNe 
(Extended Data Table 4).

\subsection{Data and Code Availability:}

The observational data analysed during the current study are available in NASA's HEASARC repository 
(https://heasarc.gsfc.nasa.gov). The atomic codes utilized in this study are also available online 
(AtomDB: http://www.atomdb.org/,  SPEX: https://www.sron.nl/astrophysics-spex).

\end{methods}

\clearpage


\begin{figure}
  \begin{center}
	\includegraphics[width=8.5cm]{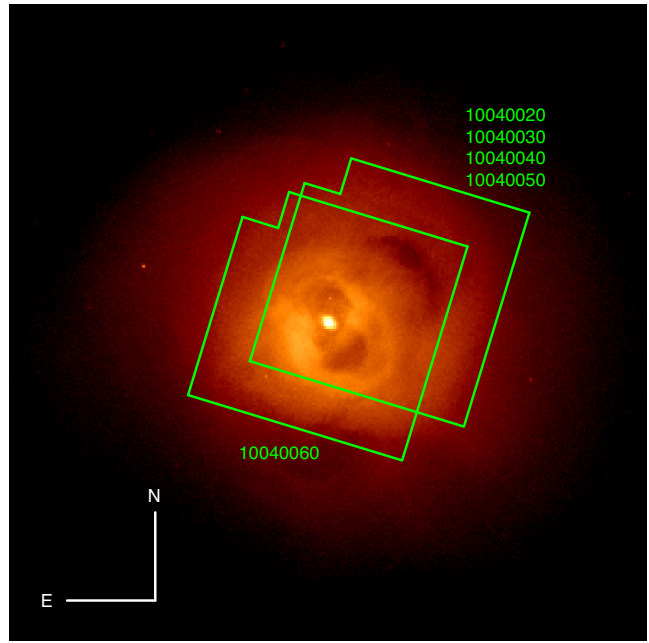}	
	\captionsetup{labelformat=empty,labelsep=none}
\caption{Extended Data Figure 1: 
{\bf The SXS FoV overlaid on a Chandra image.} 
The corresponding Sequence IDs of the Hitomi observations are given. 
Each side of the SXS has an angular size of $3'$ ($\approx$ 64\,kpc).
\label{ex1}}
  \end{center}
\end{figure}


\begin{figure}
  \begin{center}
	\includegraphics[width=16cm]{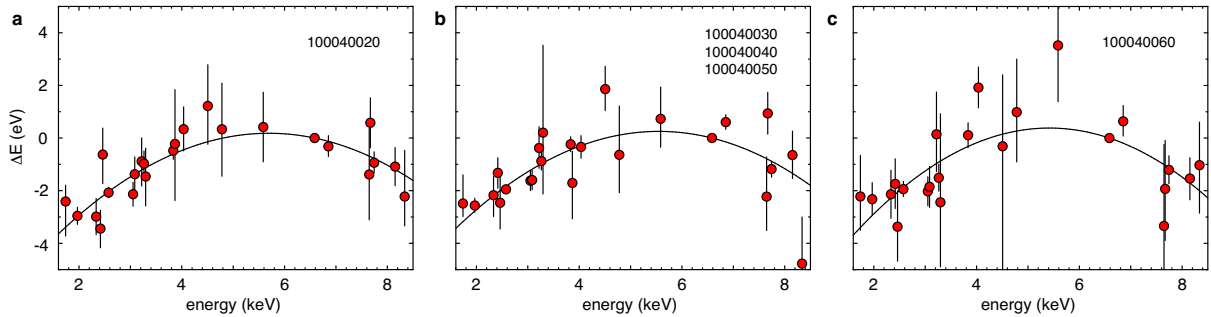}	
	\captionsetup{labelformat=empty,labelsep=none}
\caption{Extended Data Figure 2: 
{\bf Additional gain correction.}
The data points indicate the difference between the measured and theoretical energies 
($\Delta E = E' - E_0$, where $E'$ and $E_0$ are measured and theoretical energies, respectively)
of each detected line at the given X-ray energy. The best-fit parabolic functions are 
given as the solid curves. The error bars correspond to the 1$\sigma$ confidence level. 
Panels (a), (b), and (c) are the results from Sequence 100040020, 100040030--50 (combined), 
and 100040060, respectively. 
\label{ex2}}
  \end{center}
\end{figure}

\clearpage

\begin{figure}
  \begin{center}
	\includegraphics[width=16cm]{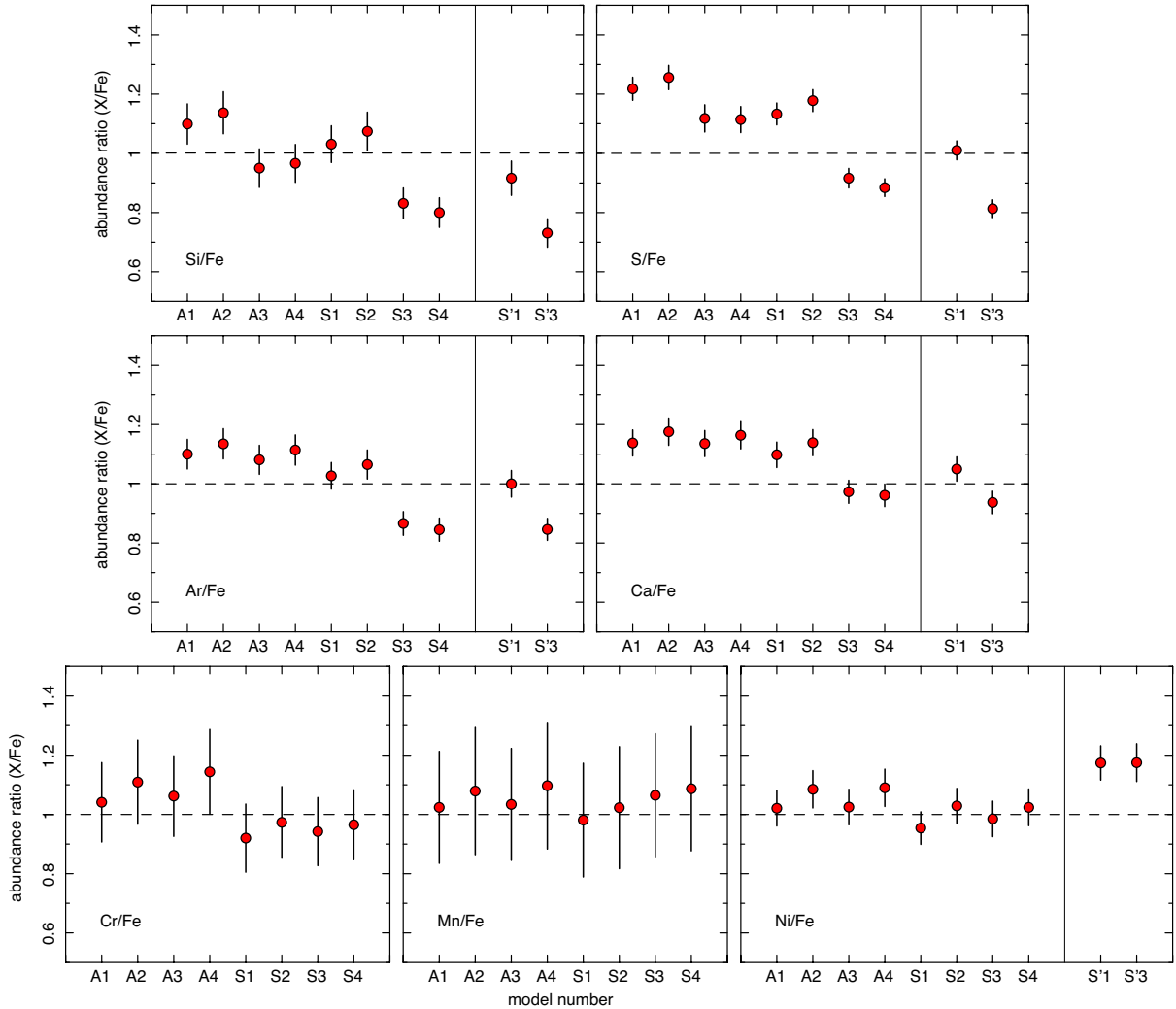}	
	\captionsetup{labelformat=empty,labelsep=none}
\caption{Extended Data Figure 3: 
{\bf Elemental abundances measured with different model assumptions.}
``A'' and ``S'' indicate the results for the atomic codes AtomDB v.3.0.8
and SPEX v.3.03, respectively; ``S$'$'' an old atomic model (SPEX v.2.05
that does not contain Cr and Mn line data). Numerical designations are
as follows.  1: one-temperature fit with the Fe XXV RS effect. 2:
one-temperature fit without the RS effect. 3: two-temperature fit with
the RS effect. 4: two-temperature fit without the RS effect. 
The error bars are at a 1$\sigma$ confidence level. 
\label{ex3}}
  \end{center}
\end{figure}

\clearpage

\begin{figure}
  \begin{center}
	\includegraphics[width=8.5cm]{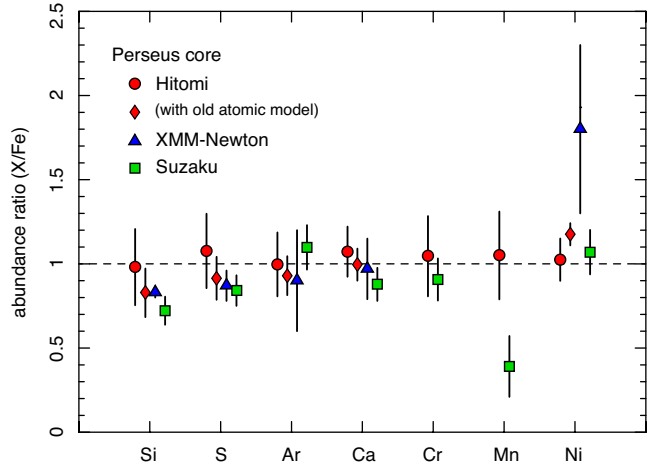}	
	\captionsetup{labelformat=empty,labelsep=none}
\caption{Extended Data Figure 4: 
{\bf Elemental abundances of the Perseus Cluster core compared among X-ray measurements.}  
The values are relative to the solar abundances\cite{Lodders09} with respect to Fe. 
The red circles are identical to those in Fig.\,2 (in main body), representing the SXS measurements 
with error bars including both 1$\sigma$ statistical uncertainty and systematic uncertainty. 
The red diamonds are the SXS measurement with an outdated atomic model 
that was used in the previous XMM-Newton results. 
The blue triangles represent the XMM-Newton results\cite{Mernier16a}, 
identical to those in Fig.\,2. 
The green squares are abundances obtained by Suzaku observations 
of the innermost $2'$ region of the Perseus Cluster\cite{Tamura09} but are 
converted relative to the updated solar abundance table\cite{Lodders09} 
for direct comparison with the other measurements. The error bars are also 
converted to the statistical uncertainty at a 1$\sigma$ confidence level.
\label{ex4}}
  \end{center}
\end{figure}

\clearpage

\begin{figure}
  \begin{center}
	\includegraphics[width=12cm]{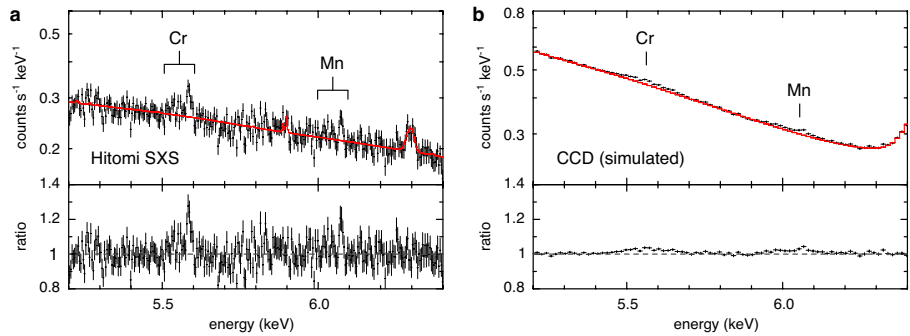}	
	\captionsetup{labelformat=empty,labelsep=none}
\caption{Extended Data Figure 5: 
{\bf Weak emission lines at different energy resolutions.}  
(a) SXS spectrum of the Perseus Cluster around the Cr and Mn emission. The red line 
is the best-fit model (Model A1) but the Cr and Mn abundances are set to zero. 
The bottom panel shows the ratio between the data and model. 
The error bars correspond to the 1$\sigma$ confidence level. 
(b) Simulated spectrum at the energy resolution of the XMM-Newton MOS1 detector
(representative of CCD data), where the best-fit model for the SXS data and 
sufficiently long exposure time (4\,Ms) are assumed.
This comparison demonstrates the robustness of our measurements of the weak emission 
lines with high resolution spectroscopy (see Methods for details).
\label{ex5}}
  \end{center}
\end{figure}

\bigskip

\begin{figure}
  \begin{center}
	\includegraphics[width=12cm]{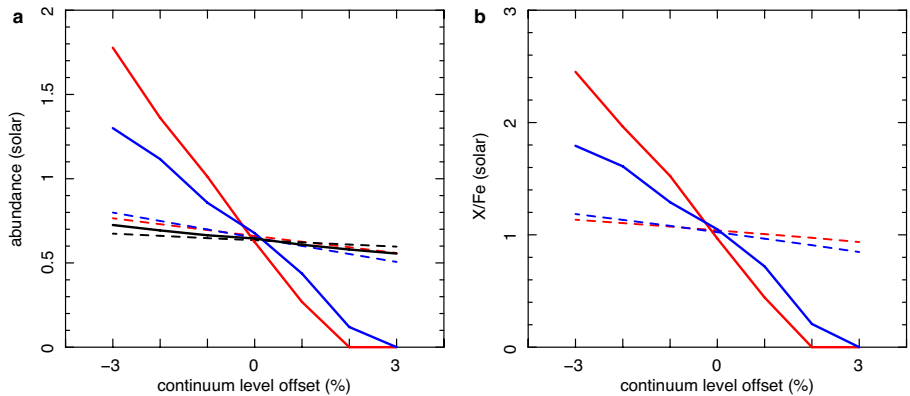}	
	\captionsetup{labelformat=empty,labelsep=none}
\caption{Extended Data Figure 6: 
{\bf Effect of potential bias in the continuum level estimate on the abundance 
measurement using weak emission lines.}  
(a) Abundances of Cr (red), Mn (blue), and Fe (black) determined by intentionally giving a small offset to 
the continuum normalization (within $\pm$\,3\% of the measured value). 
The solid and dashed lines are obtained from our test analysis of the simulated CCD spectrum (Extended Data Figure 5(b)) 
and the Hitomi spectrum, respectively. This illustrates that the CCD measurement of Cr and Mn abundances is sensitive to 
the accuracy of the continuum level determination because of the weakness of the emission and the low spectral resolution. 
The Fe abundance is less subject to such uncertainty even in the CCD measurement owing to the much larger equivalent width 
of the emission. 
(b) Abundance ratios of Cr/Fe (red) and Mn/Fe (blue) calculated using the values in panel (a) as a function of  
offset in the continuum level. 
\label{ex6}}
  \end{center}
\end{figure}

\clearpage

\begin{figure}
  \begin{center}
	\includegraphics[width=15cm]{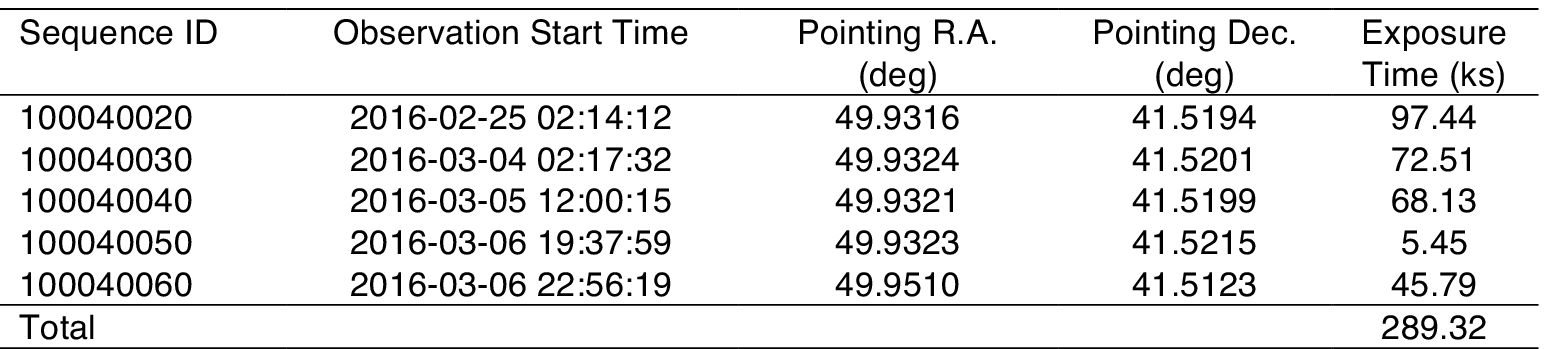}	
	\captionsetup{labelformat=empty,labelsep=none}
\caption{Extended Data Table 1: 
{\bf Summary of the observations.} \
Sequences 100040030, 40, and 50 are continuous observations, 
and separated just for the data processing reason.
\label{ext1}}
  \end{center}
\end{figure}

\bigskip
\bigskip

\begin{figure}
  \begin{center}
	\includegraphics[width=6cm]{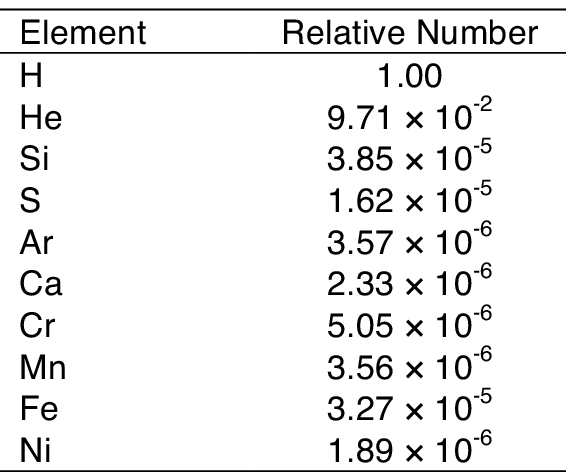}	
	\captionsetup{labelformat=empty,labelsep=none}
\caption{Extended Data Table 2: 
{\bf Solar abundance table\cite{Lodders09} referred in this work.}  
\label{ext2}}
  \end{center}
\end{figure}

\clearpage

\begin{figure}
  \begin{center}
	\includegraphics[width=12cm]{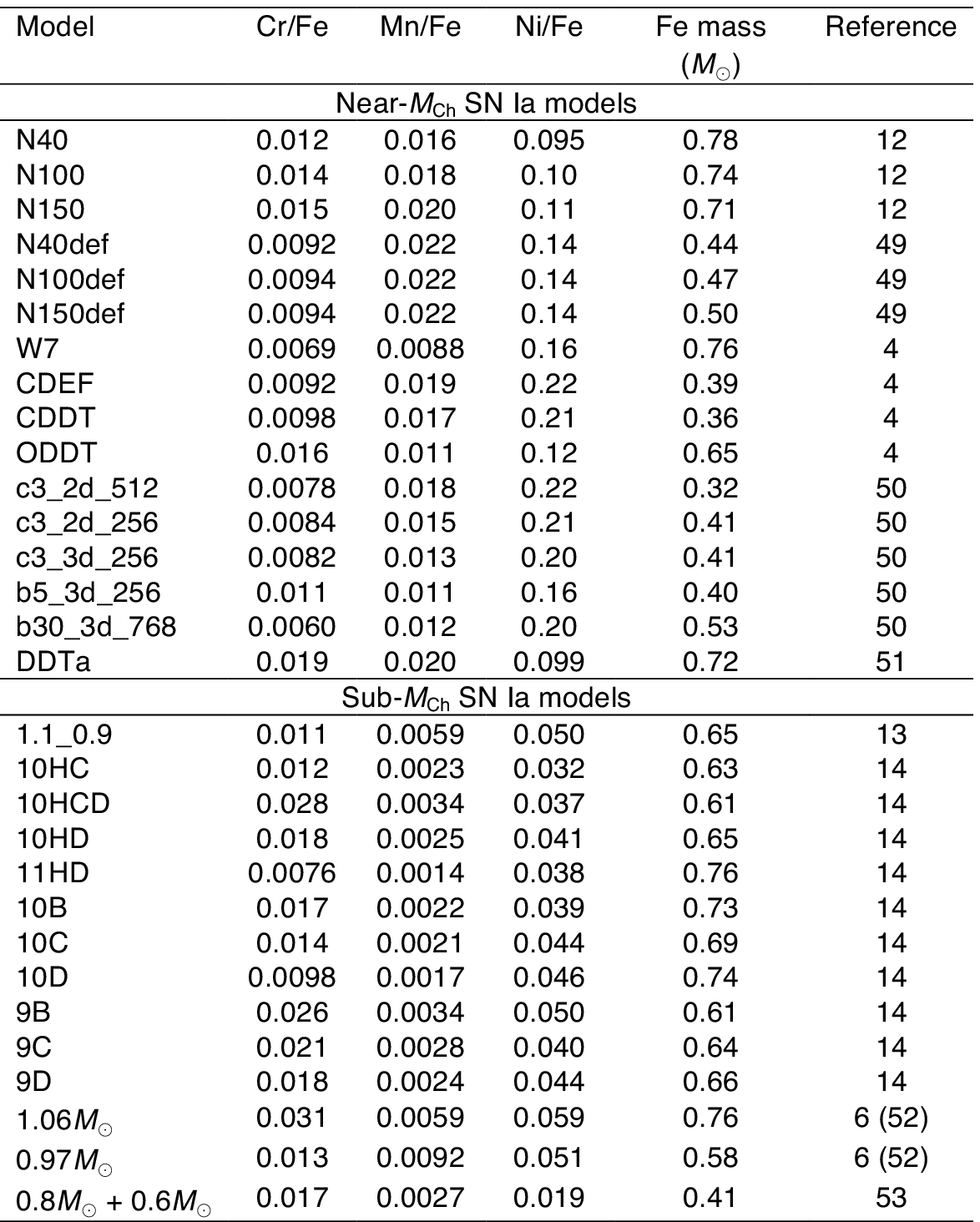}	
	\captionsetup{labelformat=empty,labelsep=none}
\caption{Extended Data Table 3: 
{\bf Mass ratios among the Fe-peak elements in SN Ia models.}  
\label{ext3}}
  \end{center}
\end{figure}

\clearpage

\begin{figure}
  \begin{center}
	\includegraphics[width=16cm]{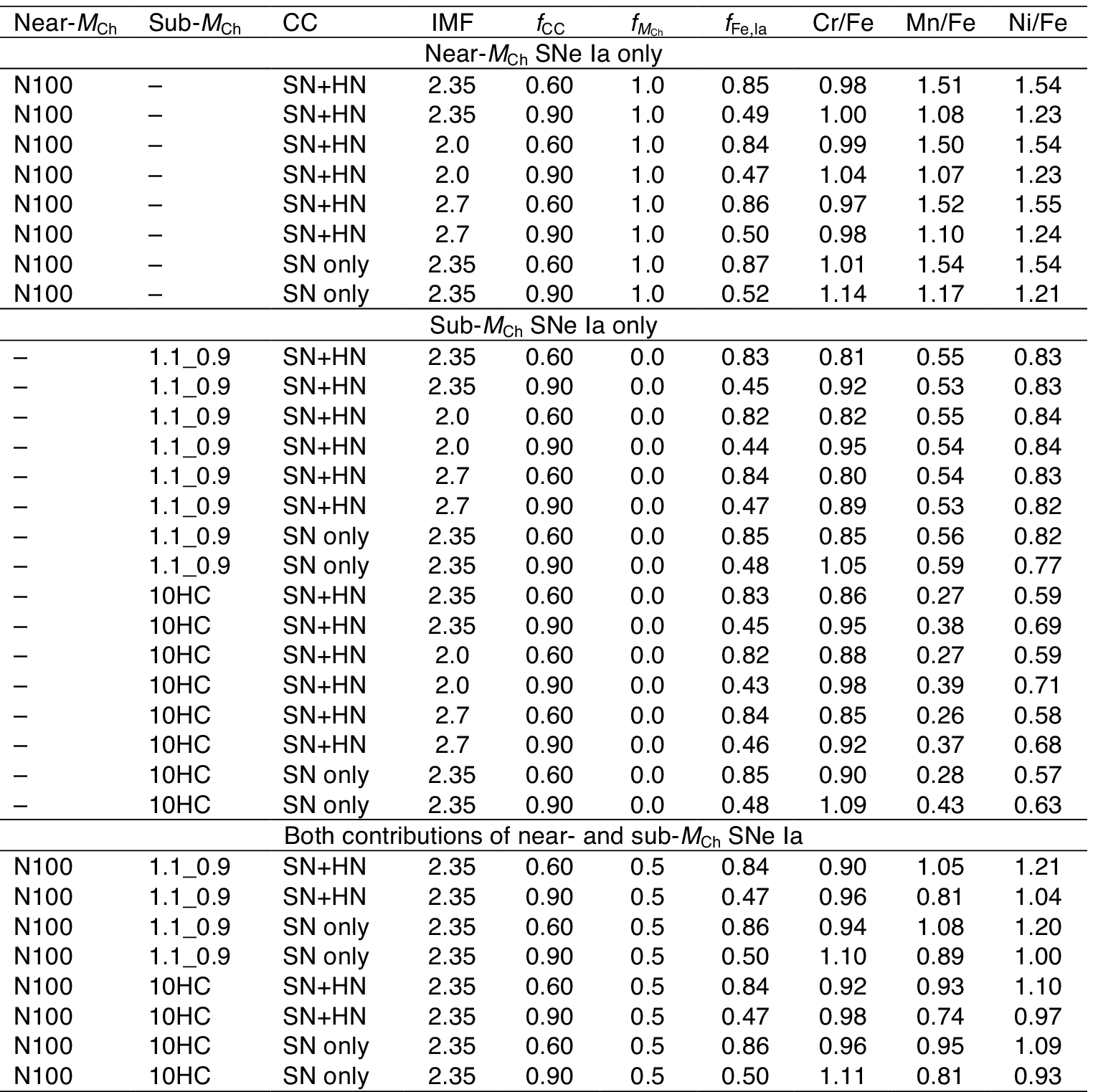}	
	\captionsetup{labelformat=empty,labelsep=none}
\caption{Extended Data Table 4: 
{\bf Example calculations of SN nucleosynthesis models for comparison with the observation.}   \
The first three columns indicate the name and/or combination of SN models, 
and the fourth column the assumed index of the IMF. 
$f_{\rm CC}$, $f_{M_{\rm ch}}$, are $f_{\rm Fe,Ia}$ 
the number fraction of CC SNe: $N_{\rm CC}/(N_{\rm Ia}+N_{\rm CC})$, 
the number fraction of near-\Mch\ SNe Ia among the total number of SNe Ia: $N_{M_{\rm ch}}/N_{\rm Ia}$, 
and the mass fraction of Fe originating from SNe Ia, respectively.
The remaining columns indicate abundance ratios relative to the solar values\cite{Lodders09}.
\label{ext4}}
  \end{center}
\end{figure}

\clearpage

\section*{Supplementary Information}

\begin{addendum}

 \item[Acknowledgements] 
H. Yamaguchi thanks Drs.\ Ivo Seitenzahl, R\"udiger Pakmor, Eduardo Bravo, and Nozomu Tominaga 
for providing the SN nucleosynthesis models used in this paper. 
We thank the support from the JSPS Core-to-Core Program.  
We acknowledge all the JAXA members who have contributed to the ASTRO-H
(Hitomi) project.  All U.S. members gratefully acknowledge support
through the NASA Science Mission Directorate. Stanford and SLAC
members acknowledge support via DoE contract to SLAC National
Accelerator Laboratory DE-AC3-76SF00515.
Part of this work was performed under the auspices of the U.S. DoE by
LLNL under Contract DE-AC52-07NA27344.
Support from the European Space Agency is gratefully
acknowledged. 
French members acknowledge support from CNES, the Centre
National d'Etudes Spatiales. 
SRON is supported by NWO, the Netherlands Organization for Scientific Research. 
Swiss team acknowledges support of the Swiss Secretariat for Education, Research and Innovation (SERI).
The Canadian Space Agency is acknowledged for the support of Canadian members.  
We acknowledge support from
JSPS/MEXT KAKENHI grant numbers 
15H00773, 15H00785, 15H02090, 15H03639, 15H05438, 15K05107, 15K17610, 15K17657, 16H00949, 16H06342, 16K05295, 16K05296, 16K05300, 16K13787, 16K17672, 16K17673, 21659292, 23340055, 23340071, 23540280, 24105007, 24540232, 25105516, 25109004, 25247028, 25287042, 25400236, 25800119, 26109506, 26220703, 26400228, 26610047, 26800102, JP15H02070, JP15H03641, JP15H03642, JP15H03642, JP15H06896, JP16H03983, JP16K05296, JP16K05309, and JP16K17667.
The following NASA grants are acknowledged: NNX15AC76G, NNX15AE16G, NNX15AK71G, NNX15AU54G, NNX15AW94G, and NNG15PP48P to Eureka Scientific.
H. Akamatsu acknowledges support of
NWO via Veni grant.  
C. Done acknowledges STFC funding under grant ST/L00075X/1.  
A. Fabian and C. Pinto acknowledge ERC Advanced Grant 340442.
P. Gandhi acknowledges JAXA International Top Young
Fellowship and UK Science and Technology Funding Council (STFC) grant
ST/J003697/2. 
Y. Ichinohe and K. Nobukawa are supported by the Research Fellow of JSPS for Young Scientists.
N. Kawai is supported by the Grant-in-Aid for Scientific Research on Innovative Areas ``New Developments in Astrophysics Through Multi-Messenger Observations of Gravitational Wave Sources''.
S. Kitamoto is partially supported by the MEXT Supported Program for the Strategic Research Foundation at Private Universities, 2014-2018.
B. McNamara and S. Safi-Harb acknowledge support from NSERC.
T. Dotani, T. Takahashi, T. Tamagawa, M. Tsujimoto and Y. Uchiyama acknowledge support from the Grant-in-Aid for Scientific Research on Innovative Areas ``Nuclear Matter in Neutron Stars Investigated by Experiments and Astronomical Observations''.
N. Werner is supported by the Lend\"ulet LP2016-11 grant from the Hungarian Academy of Sciences.
D. Wilkins is support by NASA through Einstein Fellowship grant number PF6-170160, awarded by the Chandra X-ray Center, operated by the Smithsonian Astrophysical Observatory for NASA under contract NAS8-03060
We thank contributions by many companies, including
in particular, NEC, Mitsubishi Heavy Industries, Sumitomo Heavy
Industries, and Japan Aviation Electronics Industry.

Finally, we acknowledge strong support from the following engineers.
JAXA/ISAS: Chris Baluta, Nobutaka Bando, Atsushi Harayama, Kazuyuki Hirose, Kosei Ishimura, Naoko Iwata, Taro Kawano, Shigeo Kawasaki, Kenji Minesugi, Chikara Natsukari, Hiroyuki Ogawa, Mina Ogawa, Masayuki Ohta, Tsuyoshi Okazaki, Shin-ichiro Sakai, Yasuko Shibano, Maki Shida, Takanobu Shimada, Atsushi Wada, Takahiro Yamada; JAXA/TKSC: Atsushi Okamoto, Yoichi Sato, Keisuke Shinozaki, Hiroyuki Sugita; Chubu U: Yoshiharu Namba; Ehime U: Keiji Ogi; Kochi U of Technology: Tatsuro Kosaka; Miyazaki U: Yusuke Nishioka; Nagoya U: Housei Nagano; NASA/GSFC: Thomas Bialas, Kevin Boyce, Edgar Canavan, Michael DiPirro, Mark Kimball, Candace Masters, Daniel Mcguinness, Joseph Miko, Theodore Muench, James Pontius, Peter Shirron, Cynthia Simmons, Gary Sneiderman, Tomomi Watanabe; ADNET Systems: Michael Witthoeft, Kristin Rutkowski, Robert S. Hill, Joseph Eggen; Wyle Information Systems: Andrew Sargent, Michael Dutka;
Noqsi Aerospace Ltd: John Doty; Stanford U/KIPAC: Makoto Asai, Kirk Gilmore; ESA (Netherlands): Chris Jewell; SRON: Daniel Haas, Martin Frericks, Philippe Laubert, Paul Lowes; U of Geneva: Philipp Azzarello; CSA: Alex Koujelev, Franco Moroso.

\end{addendum}

\end{document}